# Tunable ferroelectricity in artificial tri-layer superlattices comprised of non-ferroic components

K. Rogdakis[1], J.W. Seo[2,†], Z. Viskadourakis[1], Y. Wang[3], L.F.N. Ah Qune[2], E. Choi[4], J. D. Burton[3], E. Y. Tsymbal[3], J. Lee[4], and C. Panagopoulos[1, 2, 5]

[1] Electron Complexity Laboratory, Institute of Electronic Structure and Laser, Foundation for Research and Technology-Hellas, 70013 Heraklion, Greece

[2] Division of Physics and Applied Physics, Nanyang Technological University, Singapore 637371, Singapore

[3] Department of Physics and Astronomy, Nebraska Center for Materials and Nanoscience, University of Nebraska, Lincoln, Nebraska 68588-0299, USA

[4] School of Advanced Materials Science and Engineering, Sungkyunkwan University, Suwon 440-746, Republic of Korea

[5] Department of Physics, University of Crete, 71003 Heraklion, Greece

[†]Present address: School of Advanced Materials Science and Engineering, Sungkyunkwan University, Suwon 440-746, Republic of Korea.

Correspondence and requests for materials should be addressed to J. L. (email: jclee@skku.edu) or C. P. (email: christos@ntu.edu.sg).

**Heterostructured material systems devoid of ferroic components are presumed not to display ordering associated with ferroelectricity. In heterostructures composed of transition metal oxides, however, the disruption introduced by an interface can affect the balance of the competing interactions among electronic spins, charges and orbitals. This has led to the emergence of properties absent in the original building blocks of a heterostructure, including metallicity, magnetism and superconductivity. Here we report the discovery of ferroelectricity in artificial tri-layer superlattices consisting solely of non-ferroelectric $NdMnO_3/SrMnO_3/LaMnO_3$ layers. Ferroelectricity was observed below 40 K exhibiting strong tunability by superlattice periodicity. Furthermore, magnetoelectric coupling resulted in 150% magnetic modulation of the polarization. Density functional calculations indicate that broken space inversion symmetry and mixed valency, because of cationic asymmetry and interfacial polar discontinuity, respectively, give rise to the observed behavior. Our results demonstrate the engineering of asymmetric layered structures with emergent ferroelectric and magnetic field tunable functions distinct from that of normal devices, for which the components are typically ferroelectrics.**

The shortage of naturally occurring materials possessing both ferroelectricity[1,2] and magnetoelectric (ME) coupling[3-5] has propelled the search for artificial structures where individual ferroelectric (FE) and magnetic layers cause indirect enhancement of ME coupling based on interfacial exchange interactions[6,7]. Such structures may include superlattices (SLs) of FE oxide components which have been found to possess improved physical properties over homogeneous thin films[8,9], for example, the shift of the FE transition temperature ($T_c$) to higher values compared to the corresponding $T_c$ for bulk materials[10,11]. It has also been proposed that the broken space inversion symmetry at a macroscopic scale in asymmetric SLs of tri-layers in an A-B-C-A-B-C- arrangement[12] could result in a permanent built-in electric field[13,14], enhanced polarization[15] and dielectric constant[14], and large non-linear optical responses[16]. Furthermore, epitaxial strain is another useful parameter which could enhance FE properties in thin films[8,17], but may also lead to an increased polarization in tri-layer structures comprised of FE materials[15]. Most notable in the search for new materials with strong field dependence of ferroelectricity, however, is the possibility for the combination of asymmetric SL layering with epitaxial strain to encourage the formation of electrically and magnetically controllable polarization, even in the absence of FE components[18]. This strategy is of particular importance considering the lack of bulk materials comprising FE properties tunable by magnetic field[3-5].

Here we report on high quality artificial SLs with asymmetric structure composed of neither ferromagnetic nor FE perovskite manganites, yet showing distinct FE behaviour tunable by an applied magnetic field. Density functional theory (DFT) calculations indicate that the broken space inversion symmetry, because of cationic asymmetry and the presence of $Mn^{3+}/Mn^{4+}$ mixed valency originating in the interfacial polar discontinuity, give rise to the observed behavior. This discovery allows for a wide range of new FE and ME materials and paves a long sought new path in the engineering[19] of low dimensional FE devices based on manganite antiferromagnets.

**Results**

**Materials growth and characterization.** We used laser molecular-beam epitaxy to grow asymmetric tri-layer SLs composed of antiferromagnetic and non-FE manganite insulators $[(NdMnO_3)_n/(SrMnO_3)_n/(LaMnO_3)_n]_m$ on single-crystalline $SrTiO_3$ (ref. 20). Parent compounds $LaMnO_3$ and $NdMnO_3$ have $Mn^{3+}$ ions and are A-type antiferromagnets, whereas $SrMnO_3$ is a G-type antiferromagnet with $Mn^{4+}$. Structural and topographic characterizations confirmed the presence of sharp interfaces with roughness less than one unit cell and a surface roughness of ~ 2 Å (ref. 21). The high quality of the samples was clarified further using detailed bulk magnetization and neutron reflectometry studies[21]. SL capacitors were prepared with fixed total thickness of 50 nm and varying $(n, m)$ = (22, 2), (5, 8) and (2, 21); this was done in order to investigate the effect of periodicity on the polarization and the ME coupling. A schematic view of a typical capacitor fabricated for our study is shown in Fig. 1a. Bare $SrTiO_3$ capacitors were treated under conditions identical to the $SrTiO_3$ substrates on which the SLs were grown and were used to check for possible contributions of the substrate to the measured emergent properties.

**Magnetically controlled electrical polarization.** Figure 1b depicts dc polarization measurements for a SL sample with a layer period (22, 2), indicating the onset of ferroelectricity below ~ 40 K and a temperature dependence characteristic of relaxors[22]. The polarization can be reversed by alternating the polarity of the applied electric field. The polarization versus electric field ($P$-$E_a$) hysteresis loop measured at a temperature T=5 K (Fig. 1c) indicates a remnant ac polarization $P_r$ ~ 0.7 µC cm$^{-2}$ following the application of 2 kV cm$^{-1}$. The butterfly loops[15] in the capacitance-voltage (C-V) characteristics (Fig. 1d) measured at 5 K add credence to the FE switching in the SLs. Above $T_c$, both the C-V shift and the $P$-$E_a$ hysteresis loop are suppressed, indicating the intrinsic nature of the observed hysteresis (Supplementary Fig. S1). Moreover, the weak frequency dependence of the $P$-$E_a$ hysteresis (Fig. 1c) excludes contribution from extrinsic effects such as space charges, interface traps or charge leakage. Cooling the sample from room temperature and in the presence of an applied magnetic field of 6 T, hereby described as "magnetic field cooling" (MFC), shifts the polarization plateau to lower temperatures (Fig. 1b). Concurrently, the absolute value of the measured polarization (more precisely at T=5 K) is enhanced by ~ 100%.

Notably, for the SLs studied here, we have recently demonstrated that the interfaces show magnetic relaxor and spin glass-like properties[21]. A modulated magnetization of antiferromagnetic and ferromagnetic layers as a function of depth was also demonstrated using polarized neutron reflectivity[21]. Combined with measurements of bulk magnetization, it was shown that the relaxor characteristics are due to the competitive interaction between ferromagnetism present mainly at the interfaces and antiferromagnetic regions in a magnetically modulated system. Therefore, further insight to the nature of the ferroelectricity and ME coupling may be obtained from measurements of polarization (Supplementary Fig. S2) and magnetization[21] in zero field cooling conditions. The slim-loop like $P$-$E_a$ hysteresis[22], the extended tail of the polarization above $T_c$ (ref. 22) and the thermal hysteresis between zero-field-cooled and field-cooled measurements of the polarization, and magnetization[21] are indicative of a ME relaxor[21-25].

**Tunability by SL periodicity.** To investigate the effect of periodicity, we performed measurements on a sample with a layer period (5, 8) in a manner similar to the measurements made on sample (22, 2). We find the magnitude of the polarization increases compared to the (22, 2)-SL while $T_c$ shifts to lower temperatures, namely around 15 K (Fig. 2a). The ME effect is however, weaker compared to SLs with a layer period (22, 2), resulting to a less than 20% increase in the measured polarization at 5 K during MFC at 6 T (Fig. 2b). Corresponding $P$-$E_a$ hysteresis loops measured at 5 K (Supplementary Fig. S3) resulted in $P_r$ ~2.7 μC cm$^{-2}$, following the application of 2 kV cm$^{-1}$. The larger polarization compared to the (22, 2)-SLs is also apparent from the enhanced shift in the corresponding C-V characteristics (Supplementary Fig. S3).

Further reduction in periodicity namely, SLs with (2, 21) shifts $T_c$ to 25 K (Fig. 2a). Applying 6 T during MFC yields almost a 150% increase in the absolute value of the polarization (Fig. 2b). Corresponding $P$-$E_a$ hysteresis loops for different frequencies are also depicted in Supplementary Fig. S1 (see Supplementary Methods for a discussion). Although the (2, 21)-SLs exhibit a lower polarization compared to the (5, 8)-SLs, the ME coupling is significantly enhanced, exceeding that of the (22, 2)-SLs. Figure 2a depicts a comparison between the temperature dependence of the polarization for the three sets of SLs studied here, as well as for the bare treated SrTiO$_3$ substrates. Compared to the SLs, the polarization of the substrate is considerably lower, indicating the ferroelectricity and ME coupling studied here arise mostly from the SL structure grown on the SrTiO$_3$ substrate.

An overall comparison of the samples depicted in Fig. 2 reveals a strong dependence[23,26] of the electric and magnetic properties of the SLs on layer thickness and / or on the total number of tri-layers integrated within the capacitor dielectric stack. It is therefore, apparent that the interfaces play a central role in the emergent FE properties. In particular, the periodicity of the SL affects both the magnitude of the polarization and $T_c$, as well as the strength of the ME coupling - as depicted by the effect of the magnetic field on the measured polarization (Fig. 2a,b). Notably, an enhanced polarization and weaker magnetic field dependence are observed with decreasing the thickness of the layers comprising a SL, for as long as the layer thickness is kept greater or equal to five unit cells. Bulk magnetization measurements revealed a similar thickness dependence[21], with the (5, 8)-SLs exhibiting concurrently the highest magnetization (Fig. 2c). The abovementioned monotonic thickness dependence is reversed for both the polarization and the magnetization when the thickness of the layers is lower than five. An anti-correlation[23] between the thickness dependence of the ME coupling and the magnetization (and polarization) is depicted in Fig. 2b,c.

**First-principles calculations.** To explore the microscopic mechanism of the reported tunable ferroelectricity, *ab-initio* DFT calculations were performed for a periodic [(LaMnO$_3$)$_2$/(SrMnO$_3$)$_2$/(NdMnO$_3$)$_2$] SL, as represented in Fig. 3a, with in-plane strain appropriate for epitaxial growth on an SrTiO$_3$ (001) substrate (see Methods for details). Because of computational constraints, we restricted our study only to short period n=2 SLs.

The formal charge layering of the $LaMnO_3$ and $NdMnO_3$ components gives rise to a polar discontinuity at the interface of these with $SrMnO_3$, leading to charge transfer and off-centering of both the Mn and the A-site cations with respect to the interconnected lattice of O octahedral in response to macroscopic internal electric fields. This off-centering is characterized by the average out-of-plane component of the displacement between metal-cations and oxygen-anions in each atomic layer, as plotted in Fig. 3b (left axis). The inherent A-site cationic asymmetry of this tri-layer SL leads to a pattern of such displacements that does not possess space inversion symmetry, and therefore to a macroscopic polarization as can be clearly seen in the red points in Fig. 3b, which we refer to as the +$P$ state. For such a structure to be considered as FE, however, there must be a complementary state with opposite polarization orientation. Such a state can be constructed by inverting the pattern of atomic displacements, but keeping the chemical identity of each site the same as in the +$P$ state. After relaxation, we find a pattern of metal-oxygen displacements approximately inverted with respect to the +$P$ state. We label this inverted state as –$P$ and plot the displacements in blue in Fig. 3b. Given these two stable polarization states with opposite orientation, we can estimate the energy profile associated with polarization reversal by linearly scaling the atomic structure between the two stable minima, keeping track of the total energy of each tested structure. This scaling is parameterized by the dimensionless parameter $\lambda$, where $\lambda = \pm 1$ corresponds to the $\pm P$ state. The -$P$ state has energy 50 meV/cell above the +$P$ state. The resulting double-well energy profile, plotted in Fig. 3c, is entirely consistent with what is expected for ferroelectricity in compositionally ordered structures with broken inversion symmetry[13].

The bi-stability of the structure is intimately coupled to the charge transferred in response to the interfacial polar discontinuity. Figure 3d shows the Mn-3$d$ projected local density of states (LDOS) near the Fermi level ($E_F$) on each of the MnO$_2$ atomic layers in the +$P$ state (corresponding plots for the –$P$ state are shown in Fig. 3e). Although all three components of the SL are insulating in the bulk (see the filled plots in the background of the second, fourth and sixth panels), there is non-zero LDOS at $E_F$ for the "up"-spin electrons on the MnO$_2$ layers above and below the central G-type ordered layer. All other MnO$_2$ layers are essentially gapped, except the layer sandwiched on both sides by LaO layers, where there is non-zero LDOS at $E_F$ for the "down"-spin states. The reverse polarization state (Fig. 3e), –$P$, exhibits similar charge transfer profile, except in this case the "down"-spin charge is trapped in the NdMnO$_3$ layer (compare bottom layers in Fig. 3d,e), whereas the LaMnO$_3$ layer, which held the down-spin charge in the +$P$ state is now essentially bulk-like (compare the second layer from the top in Fig. 3d,e). The difference in the "down"-spin charged states at the LaMnO$_3$ and NdMnO$_3$ layers between +$P$ and –$P$ states, respectively, is essential to produce the polarization bi-stability. This bi-stability in the position of the interface charge clearly gives rise to the switchable behavior of the structure similar to that known for FE thin films sandwiched between two metal electrodes, and the interface charge may be interpreted as the "screening" charge for polar displacements in the heterostructure[27].

Furthermore, there is a strong coupling between structural, electronic and magnetic degrees of freedom in the heterostructure. This can be seen from Fig. 3b (right axis), which shows the difference in the layer-by-layer total magnetic moment (per Mn atom) between the +$P$ and –$P$ states, $\Delta M$. The change in the magnetic moment is entirely consistent with the hopping of "down"-spin electrons from the LaMnO$_3$ layer to the NdMnO$_3$ layer. The ME origin of the switchable polarization in the SLs suggests that changing magnetic moment alignment, for example, by applying an external magnetic field will strongly influence the charge transfer and thus leads to the magnetic modulation of the polarization as it is observed experimentally. In particular, the application of a magnetic field will break the inversion symmetry of the magnetic ordering and will therefore favor further charge transfer to one layer over the other, and therefore lead to a larger polarization.

Our calculations suggest that the ferromagnetism originates from the interfaces, consistent with ref. 21. On the other hand, the relaxor behavior[21] could not be seen from the theoretical results, which describe a static and zero-temperature behavior revealing thus only the ground state of the system. Nevertheless, our DFT results indicate that the magnetic order and the origin of ferroelectricity are coupled (through the "trapping" and "hopping" of "down"-spin charge), therefore, ferroelectricity should have relaxor character too (as depicted in Fig. 1). Finally, we note that as the polar displacements as well as the largest changes in the magnetization profile are localized at the interfaces (Fig. 3b), the net polarization and magnetization should scale with the density of interfaces, that is, a $1/n$ dependence on the lattice periodicity, $n$, which is consistent with the experimental thickness dependence depicted in Fig. 2c for $n \geq 5$.

**Discussion**

The SLs studied here are comprised of three manganite perovskites each having a different A-site ion size thus introducing modulated strain and effective chemical doping[19,26]. Reciprocal mapping using synchrotron x-ray diffraction showed, that although the in-plane lattice constants of all films match the lattice constant of the substrate, our SLs exhibit a tetragonal lattice distortion. Even though our experiments do not allow an estimate of atomic displacement, structural distortions and chemical engineering including Mn off-centering have been suggested to encourage FE instabilities in various perovskite systems[28,29]. Besides, electronic reconstruction arising from the interfacial $Mn^{3+}$ / $Mn^{4+}$ mixed valency[30-32] would favor the formation of local electric dipolar moments accommodated with a magnetic degree of freedom. Remarkably, our *ab-initio* calculations revealed that a synergy between asymmetric lattice distortion and atomic displacement, mixed valency and tunability of charge itinerancy by an applied magnetic field finally leads to a magnetically controlled FE polarization. Our discovery is a long sought[19] demonstration for emergent ferroelectricity and magnetoelectricity in an artificial system originating purely in the broken space inversion symmetry as well as in the strong carriers' complexity and correlations at the nanoscale. A plethora of artificial FE and ME low dimensional devices may now be designed and fabricated following the recipe for a tri-layer SL comprised of transition metal oxides incorporating mixed valency, compatible magnetic ordering and asymmetric cation displacement.

In conclusion, we demonstrated the emergence of ferroelectricity in artificial SLs composed of materials that are non-FE in their bulk form. We also observed up to 150% increase in the measured *P* by the application of a magnetic field. Engineering systems with tunable ferroelectricity and ME coupling in asymmetrically layered devices comprised of materials abundant in synthetic chemistry such as those used in the present study would widen the versatility of electric polarization functionality.

## Methods

**Capacitor fabrication and general measurement details.** Circular metallic (Pt/Au) pads of various diameters (100 - 300 μm) were patterned on the SLs using electron-beam evaporation. The thickness of the $SrTiO_3$ substrate was ~ 500 μm. A metallic layer was deposited on the underside of the substrate using the same evaporation technique as for the top electrodes and served as a common bottom capacitor plate. Silver paint contacts were placed on the top and bottom metallic pads of the samples for both polarization and capacitance measurements. A systematic error of < 5 % was obtained when repeating measurements several times under identical experimental conditions. The results were accurately reproducible using many different capacitors of the same sample type. No aging effect was observed.

**Capacitance - voltage measurements.** The capacitance – voltage (C-V) measurements were performed at various temperatures using an LCR meter with an AC signal frequency and magnitude of 20 kHz and 200 mV, respectively. Measurements using different voltage scanning directions (defined by arrows in the corresponding plots) revealed the FE switching of the SLs. The virgin curves of voltage scanning from 0 V to maximum value are not shown.

**DC polarization data processing.** Polarization ($P$) was measured using a pyroelectric current technique with a bias electric field ($E_a$) applied perpendicular to the plane of the SL (using a Keithley electrometer 6517B). During an electric field cooling process, the samples were first polarized by applying the desired $E_a$ at high temperature and then cooled to 5 K while $E_a$ was maintained at the same bias. During the zero electric field cooling process, the samples were cooled to 5 K in zero $E_a$. The pyroelectric current and the sample temperature were then recorded as a function of time, while the sample was warmed at a constant $E_a$ and a heating rate of 5 K min$^{-1}$. A typical value of $E_a$= +100 V cm$^{-1}$ was used. (All measurements of $P$ were performed during an electric field heating process, unless otherwise stated.) For the zero magnetic field cooling (ZMFC) process, the sample was cooled to base temperature in the absence of an applied magnetic field, whereas in the case of MFC a constant magnetic field was applied during cooling. In both cases, the heating cycle was performed in the presence of a magnetic field - applied parallel to the plane of the film, unless otherwise stated. Similar tests have been reported for the magnetization of our SLs[21].

**AC polarization measurements.** Polarization versus electric field ($P$-$E_a$) hysteresis loops were measured at various temperatures using a TF 2000 analyzer with $E_a$ ranging from -2 kV cm$^{-1}$ to 2 kV cm$^{-1}$ (the thickness of the SrTiO$_3$ substrate limited the maximum applied electric field). The AC signal frequency ranged from 10 Hz to 1 kHz. An initial triangular excitation pulse was applied to establish a polarization state. Three consecutive excitation pulses followed, separated by a relaxation time of 1 s. During each pulse, the current was measured and the polarization extracted by integrating the current with respect to time.

*Ab-initio* **calculations.** Electronic and atomic structure calculations were performed within DFT applied to (LaMnO$_3$)$_2$/(SrMnO$_3$)$_2$/(NdMnO$_3$)$_2$ SLs. The calculations employed the projector augmented wave method as implemented in the Vienna Ab Initio Simulation Package code[33]. The exchange-correlation effects were treated within generalized gradient approximation (GGA). The calculations were carried out using a plane-wave basis set with a cutoff energy of 520 eV and an 8×8×4 mesh of k points in the irreducible Brillouin zone with energy converged to 0.01 meV/cell. Atomic relaxations were performed until the Hellmann-Feynman forces on atoms have become less than 30 meV/Å. The in-plane lattice constant is constrained to the experimental bulk lattice constant of SrTiO$_3$, $a$ = 3.905Å. On-site electron-electron Coulomb interactions of the Mn-3$d$ states were treated within the GGA+U approach[34]. We impose $U$ = 8.0 eV and $J$ = 1eV on Mn-3$d$ states[35,36] giving reasonable band gaps for all three components: $E_g$ = 0.99 eV, 1.2 eV and 0.23 eV for LaMnO$_3$, NdMnO$_3$ and SrMnO$_3$ respectively, which are consistent with experimental values 1.0 eV (ref. 37), 1.7 eV (ref. 38), 0.3 eV (ref. 39). Since the La-4$f$ bands lie at higher energy than that predicted by GGA, the value of $U$ for the La 4$f$ states was taken to be 11.0 eV to avoid spurious mixing with the conduction bands. The structural model was initially constructed from the bulk structures of each manganite component, all of which are found to have the tilts and rotations of oxygen octahedrons consistent with the Pnma space group. The Mn magnetic moments are arranged in a collinear fashion according to the known antiferromagnetic ordering in the bulk compounds: G-type in SrMnO$_3$ and A-type in both LaMnO$_3$ and NdMnO$_3$, as indicated in Fig. 3a. Those MnO$_2$ layers sandwiched between SrO and LaO or NdO layers continue the neighboring A-type order, while the MnO$_2$ layer sandwiched between two SrO takes on a G-type checkerboard pattern. Several other collinear magnetic order configurations were tested and were found to have higher energy.

**Acknowledgements**

We thank C. Batista and I. Martin for insightful discussions. This work was supported by the National Research Foundation, Singapore through a Competitive Research Program grant, the European Union through EURYI and MEXT-CT-2006-039047 grants. Research at the University of Nebraska-Lincoln (UNL) was supported by the National Science Foundation through Nebraska EPSCoR (Grant No. EPS-1010674) and Nebraska MRSEC (Grant No. DMR-0820521). Computations were performed at the UNL Holland Computing Center. The work at Sungkyunkwan University was supported by the National Research Foundation of Korea through Basic Science Research Program (2009-0092809).

**Author contributions**

K. R. and J. W. S. equally contributed to this work. K. R., J. W. S., J. L. and C. P. conceived and designed the experiment. K. R. carried out the polarization and capacitance measurements, interpreted the data and prepared the manuscript. J. W. S. and L. AQ contributed to the capacitance measurements and writing of the manuscript. Z. V. built the polarization experimental probe and contributed to the writing of the manuscript. Y. W. performed the ab-initio calculations and J. D. B. and E. Y. T. developed the theory portions of the manuscript. E. C. and J. L. performed the SL growth and fabricated test structures for capacitance measurements. C. P. supervised the study and co-wrote the paper. All authors discussed the results and commented on the manuscript.

**Figure Legends**

**Figure 1 │ Evidence of ferroelectric behavior in a superlattice (SL) with layer period of (22, 2).** (**a**) Schematic view of a capacitor fabricated on SLs. (**b**) Temperature ($T$) dependence of polarization ($P$) measured using the pyroelectric technique for a typical electric field ($E_a$) of +100 V cm$^{-1}$ (black curve) and -100 V cm$^{-1}$ (red curve) applied perpendicular to the plane of the SL. The $T$-dependence of $P$ with the magnetic field $H$=6 T applied parallel to the plane of the SL in magnetic field cooling conditions is also depicted (green curve). Measurements were performed in electric field cooling conditions. (**c**) $P$-$E_a$ hysteresis loops measured at T=5 K from 10 Hz to 1 kHz. (**d**) Capacitance- voltage (C-V) "butterfly"-like characteristic loop measured at T=5 K. Different voltage scanning directions are indicated by arrows.

**Figure 2 │ Evolution of ferroelectricity and magnetoelectric coupling with layer periodicity.** (**a**) Direct comparison between the temperature dependence of the measured polarization ($P$) for (5, 8), (2, 21) and (22, 2) superlattices (SLs), and bare treated SrTiO$_3$ capacitors. Solid lines represent data in zero magnetic field ($H$) and dotted lines data measured in magnetic field cooling (MFC) conditions at $H$=6 T. Electric field cooling conditions with $E_a$= +100 V cm$^{-1}$ were followed for the abovementioned $P$ measurements. (**b**) Normalized relative change of $P$ ($E_a$= +100 V cm$^{-1}$) in MFC at $H$=6 T for all SLs. (**c**) Comparison of the thickness dependence of $P$ (left axis) and magnetization (right axis) for various SLs at T=5 K. Dotted lines are a guide to the eye. The red arrow indicates the y-axis for the data in red open circles and the black arrow the y-axis for the data in black closed circles.

**Figure 3 | Results of first-principles calculations of switchable electric polarization in a (LaMnO$_3$)$_2$/(SrMnO$_3$)$_2$/(NdMnO$_3$)$_2$ superlattice (SL).** (**a**) Structure of the SL unit cell used in the calculations. Mn ions are surrounded by a distorted octahedron of O (small red spheres), and the black arrows represent the magnetic moment orientation of each Mn site. The *A*-site cations La, Sr and Nd are represented by blue, green and yellow spheres, respectively. (**b**) Average cation-anion (metal-oxygen, M-O) displacements along the SL direction (left axis). The red and blue circles correspond to the +*P* and −*P* states, respectively. The difference in average total magnetic moment between +*P* and −*P* states, Δ*M*, (right vertical axis) of each MnO$_2$ layer is also plotted (open circles). (**c**) Energy per cell as a function of the parameter *λ* which scales between the +*P* and −*P* states (see text). (**d**) Spin-polarized local density of states (LDOS) projected on the Mn-3*d* orbitals, averaged over the two inequivalent Mn sites in each MnO$_2$ layer. Each panel from top to bottom corresponds to the atomic layer depicted immediately to the left in **a**. The upper sub-panels are for up-spin electrons, and the lower sub-panels are for down-spin electrons. The filled plots in the background of the second, fourth and sixth panels are the Mn-3*d* projected LDOS in bulk LaMnO$_3$, SrMnO$_3$ and NdMnO$_3$, respectively. (**e**) Spin-polarized LDOS for the −*P* state, ordered the same as in **d**.

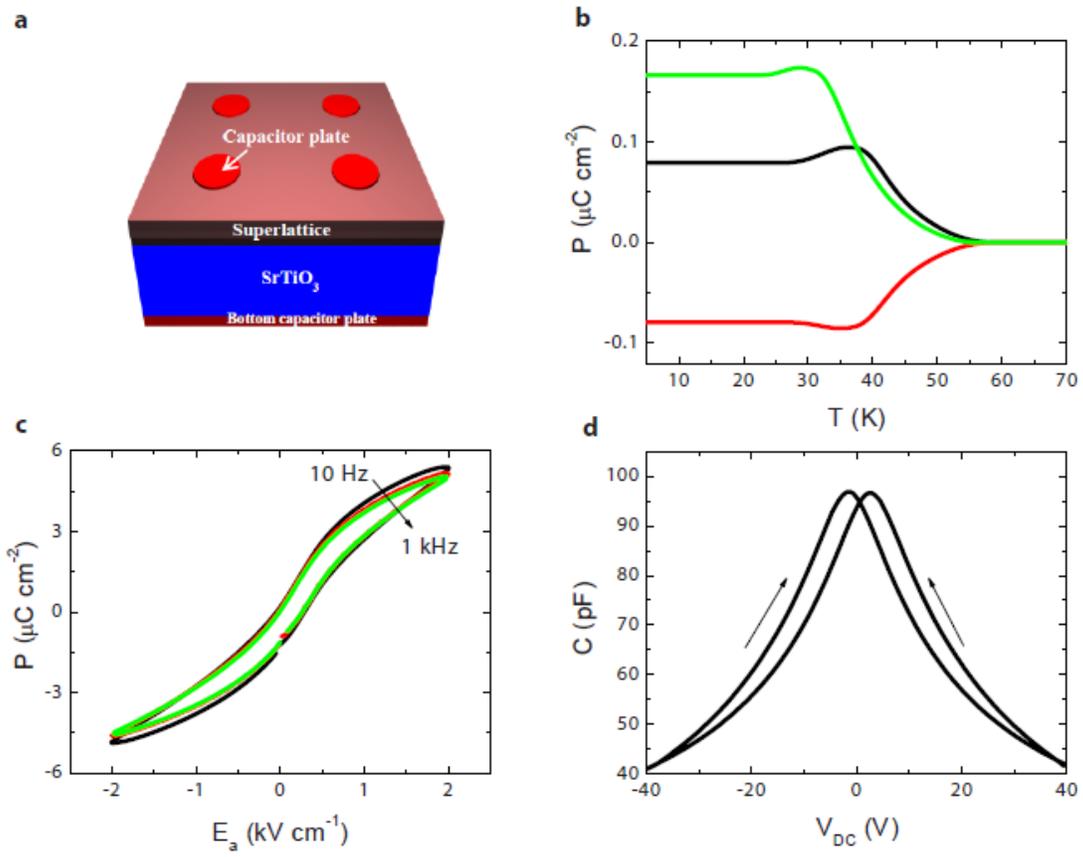

**Figure 1**

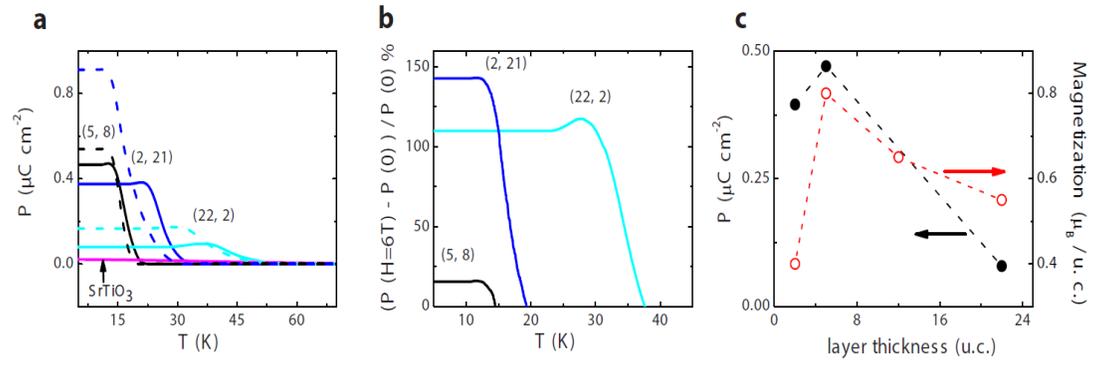

**Figure 2**

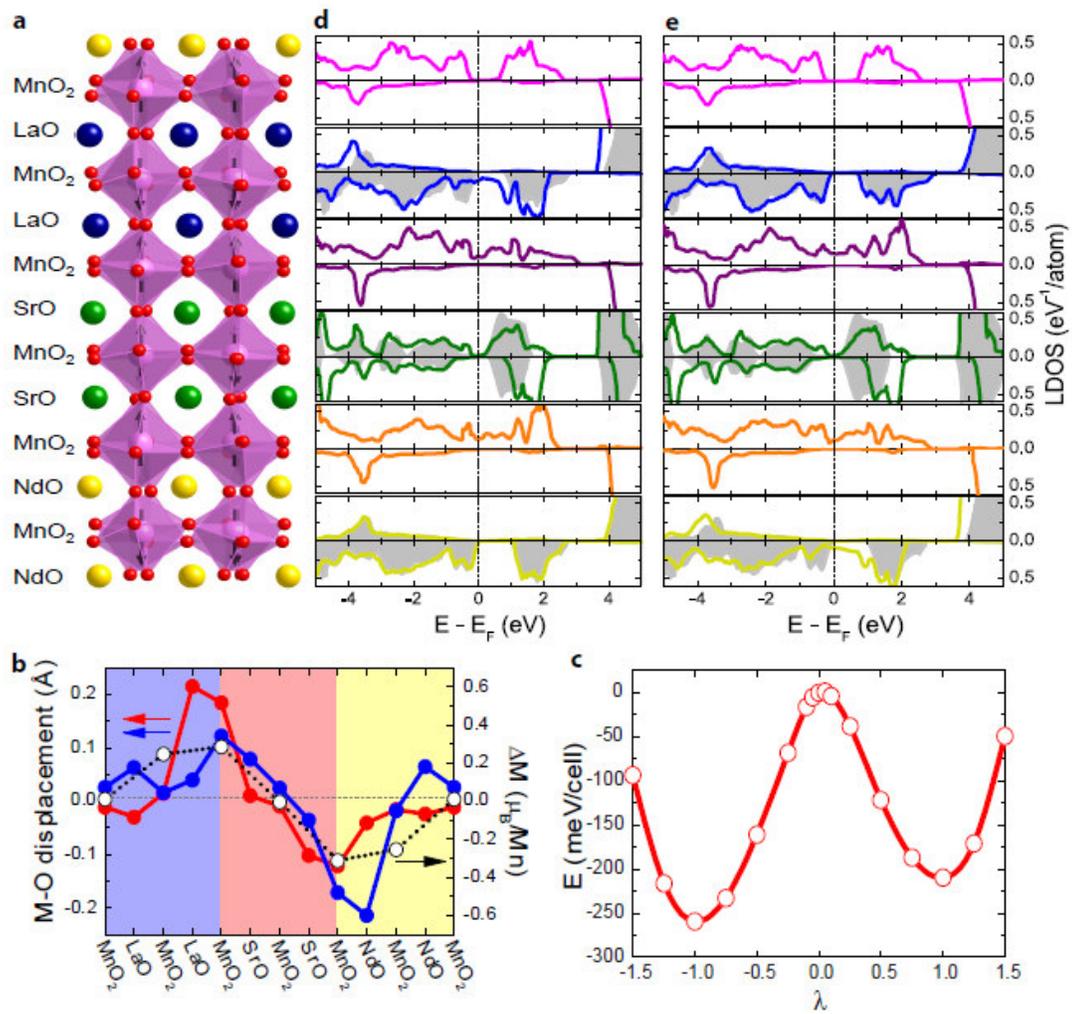

**Figure 3**

**Supplementary Figures**

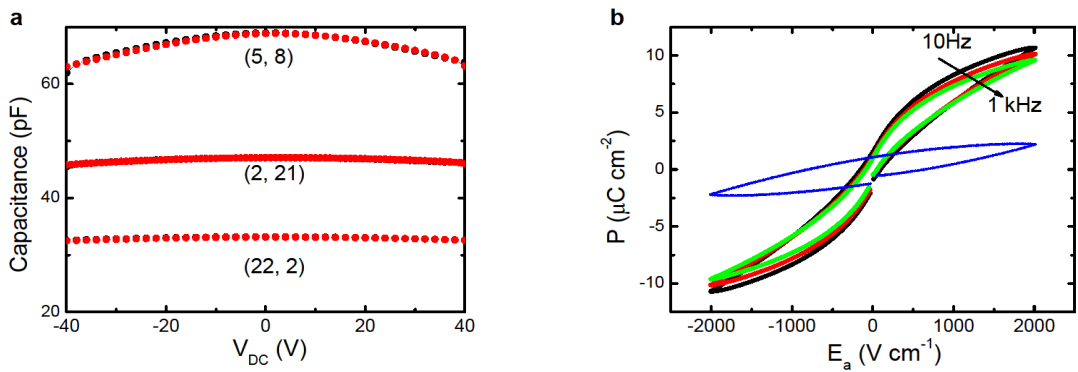

**Supplementary Figure S1 │ Evidence for suppression of hysteretic effects above the ferroelectric transition temperature.** (**a**) Capacitance-voltage (C-V) characteristics of various superlattices (SLs) measured at temperatures above their corresponding ferroelectric transition temperature ($T_c$). Data for (5, 8), (2, 21) and (22, 2) SLs have been measured at 20 K, 25 K, and 50 K, respectively. No hysteresis between different voltage scanning directions is observed. Different scanning directions are depicted in black and red color. (**b**) Polarization versus electric-field ($P$-$E_a$) loops for a SL with period (2, 21) above and below $T_c$. The $S$-like $P$-$E_a$ loop observed at low temperatures (black, red and green curves measured at 5 K in 10 Hz, 100 Hz and 1 kHz, respectively) vanishes when the sample temperature exceeds $T_c$ (blue curve measured at 30 K in 1 kHz).

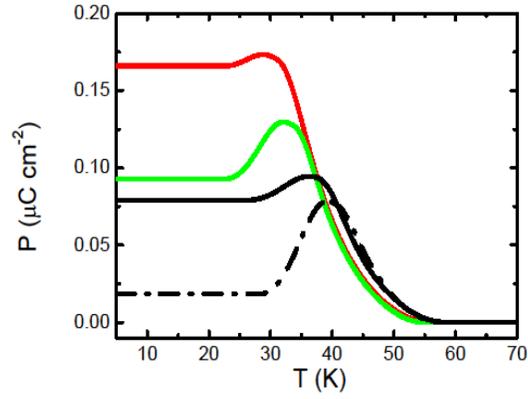

**Supplementary Figure S2 │ Observation of thermal hysteresis in polarization measurements in zero field cooling conditions.** Comparison between the temperature dependence of polarization in superlattices (SLs) with a layer period of (22, 2) in the absence of an applied magnetic field ($H$) under (i) electric field cooling (EFC) (black solid curve) and (ii) zero electric field cooling (ZEFC) (black dashed curve) conditions, and with $H$=6 T applied parallel to the plane of the SL under (iii) zero magnetic field cooling (ZMFC) (green curve) and (iv) magnetic field cooling (MFC) (red curve) conditions. Both ZMFC and MFC measurements were performed in EFC.

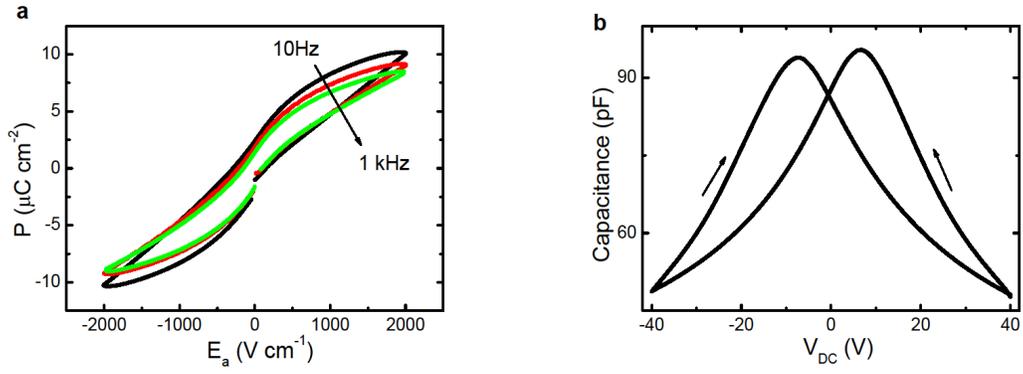

**Supplementary Figure S3 │ Hysteretic effects in a superlattice with layer period of (5, 8).** (**a**) Polarization versus electric-field (*P-E$_a$*) hysteresis loops measured at T=5 K from 10 Hz to 1 kHz. (**b**) Capacitance-voltage (C-V) "butterfly"-like characteristic loop measured at T=5 K. Different voltage scanning directions are indicated by arrows.

## Supplementary Methods

**Extrinsic contribution in AC measurements.** To study the ferroelectric (FE) characteristics of the superlattices (SLs) we employed different measurement techniques. This combination of measurements allows us to dispel possible extrinsic contributions[1,40,41]. Supplementary Fig. S1a depicts capacitance-voltage (C-V) loops for SLs measured above their corresponding FE transition temperature ($T_c$). The suppression of the shift in C-V above $T_c$ indicates that such a shift is related to FE switching in the SLs. If extrinsic effects were at play, the C-V shift would have increased with increasing temperature. Furthermore, extrinsic contributions due to charge mobility and injection, or ion motion would have dominated at higher temperatures[42], and not in the low temperature regime where ferroelectricity emerges in our samples. A charge-voltage (Q-V) hysteresis might also arise due to extrinsic effects. However, a C-V loop would depict the characteristic butterfly shape only when the material is FE (ref. 1). Furthermore, if space charges close to the electrical contacts contributed to our measurements, we would have observed a strong asymmetry and multiple peaks in the C-V plots[43]. The smooth butterfly-loops presented in the Article dismiss such a possibility. Supplementary Fig. S1b shows a corresponding polarization versus electric-field ($P$-$E_a$) hysteresis loop of a SL with a layer period of (2, 21) measured below and above the FE $T_c$. The $S$-like $P$-$E_a$ loop observed at low temperature vanishes when the sample temperature exceeds $T_c$. Furthermore, if extrinsic contributions were at play, increasing the frequency of the AC signal would result in a large difference in both the magnitude of $P$ and the shape of the $P$-$E_a$ loop[1,40,41,43]. In particular, the shape of the hysteresis loop would substantially change with frequency, with the polarization saturation vanishing at high frequency[41]. The measured polarization at 2 kV cm$^{-1}$ decreases slightly with increasing frequency and is accompanied by a weak suppression of the coercive field $E_c$. This behavior indicates that the electrical dipoles do not respond fully to the applied field at high frequencies[44].

# Supplementary References